\documentclass{aa}  

\usepackage{booktabs}

\usepackage{graphicx}
\usepackage{natbib}
\usepackage{txfonts}
\usepackage{threeparttable}
\usepackage{siunitx}
\usepackage{color}
\usepackage{xcolor}
\definecolor{mhi}{rgb}{0.6,0,0.6}

\definecolor{msp}{rgb}{0.0,0.6,0.6}

\newcommand{\om}[1]{{\color{black} #1}}

\begin{document}

   \title{The coherent motion of Cen\,A dwarf satellite galaxies remains a challenge for $\Lambda$CDM cosmology\thanks{Based on observations collected at the European Organisation for Astronomical Research in the Southern Hemisphere under ESO programmes 0101.A-0193(A) and 0101.A-0193(B).}}

\titlerunning{The co-moving plane of satellites around Centaurus\,A}
   \author{
          Oliver M\"uller\inst{1}
                              \and
        Marcel S. Pawlowski\inst{2}
         \and
        Federico Lelli\inst{3}
         \and
          Katja Fahrion\inst{4}
         \and
         Marina Rejkuba\inst{4}
         \and
         Michael Hilker\inst{4}
         \and
        Jamie Kanehisa\inst{2,}\inst{5}
         \and
        Noam Libeskind\inst{2}
         \and
         Helmut Jerjen\inst{6}
          }
          
           \institute{Observatoire Astronomique de Strasbourg  (ObAS),
Universite de Strasbourg - CNRS, UMR 7550 Strasbourg, France\\
 \email{oliver.muller@astro.unistra.fr}
\and
Leibniz-Institut fur Astrophysik Potsdam (AIP), An der Sternwarte 16, D-14482 Potsdam, Germany
\and
School of Physics and Astronomy, Cardiff University, Queens Buildings, The Parade, Cardiff,CF24 3AA, UK
\and
 European Southern Observatory, Karl-Schwarzschild Strasse 2, 85748, Garching, Germany
 \and
 Department of Physics, University of Surrey, Guildford GU2 7XH, UK
\and
 Research School of Astronomy and Astrophysics, Australian National University, Canberra, ACT 2611, Australia
}

   \date{}

% \abstract{}{}{}{}{} 
% 5 {} token are mandatory
   \abstract{The plane-of-satellites problem is one of the most severe small-scale challenges for the standard $\Lambda$CDM cosmological model:
   several dwarf galaxies around the Milky Way and Andromeda co-orbit in thin, planar structures. A similar case has been identified around the nearby elliptical galaxy \om{Centaurus\,A (Cen\,A}).
   In this Letter, we study the satellite system of Cen\,A adding twelve new galaxies with line-of-sight velocities from VLT/MUSE observations. We find 21 out of 28 dwarf galaxies \om{with measured velocities} share a coherent motion.
   Similarly flattened {and} coherently moving %systems 
   {structures} are found only in 0.2$\%$ of Cen\,A analogs in the Illustris-TNG100 cosmological simulation, independently of whether we use its dark-matter-only or hydrodynamical run. These analogs are not co-orbiting, and arise only by chance {projection}, thus they are short-lived structures in such simulations. Our findings indicate that the observed {co-rotating} planes of satellites are a persistent challenge for $\Lambda$CDM, which is largely independent from baryon physics. 
   }

   \maketitle
%
%-------------------------------------------------------------------

\section{Introduction}
One of the 
{main} challenges for {current} models of galaxy formation is the plane-of-satellites problem \citep{2018MPLA...3330004P}.  The plane-of-satellite problem has {its} roots in the mid-70s \citep{1976RGOB..182..241K,1976MNRAS.174..695L}, even though the implications for cosmology  had not been realized at that time. The {then-known} six satellites of the Milky Way (MW)
were found to be arranged in a thin, planar structure and were thought to have a tidal origin \citep{1982Obs...102..202L}. In the beginning of the new millennium, several {high-resolution} cosmological simulations became
available (e.g. \citealt{2005Natur.435..629S}), which allowed {investigating cosmological predictions}
on the scale of galaxy groups. \citet{2005A&A...431..517K} pointed out that the flattened spatial distribution of the then-known eleven Milky Way satellites 
-- a feature which was basically discovered 40 years earlier -- 
%FL: Repetition. Drop? 
is incompatible with the standard $\Lambda$ Cold Dark Matter ($\Lambda$CDM) model of structure formation, which predicts close to isotropic satellite distributions.
This conclusion was met with several rebuttals \citep[e.g., ][]{2005ApJ...629..219Z,2005MNRAS.363..146L,2009MNRAS.399..550L} showing that a certain degree of anisotropy is imprinted in the galaxy distribution from the accretion through the cosmic web \citep{2016ApJ...830..121L}, thus some galaxies {may occasionally} host flattened satellite structures. The discovery of another flattened structure of satellite galaxies around the Andromeda galaxy \citep{2006AJ....131.1405K,2006MNRAS.365..902M}
showed that these {planar distributions} {may not be} {rare} exceptions.

A major step in {understanding these structures} {was the measurement of} the {proper} motions of the classical {MW} satellites 
with the Hubble Space Telescope \citep[HST, e.g. ][]{2004AJ....128..687D,2017ApJ...849...93S}, %2007AJ....133..818P
{revealing} that {these galaxies} are also kinematically correlated \citep{2008ApJ...680..287M,2012MNRAS.423.1109P}.
{A similar kinematic} coherence was later found for the Andromeda system by measuring the
line-of-sight velocities of the dwarf galaxies \citep{2013Natur.493...62I}. Novel proper motion measurements of two satellites of the Andromeda galaxy indicate {that they, too, co-orbit along their satellite plane} \citep{2020arXiv200806055S}.

{Nowadays the existence of co-rotating satellite systems represents one of the major controversies in near-field cosmology}
\citep{2014ApJ...784L...6I,2015MNRAS.449.2576C,2015ApJ...815...19P,2016MNRAS.457.1931S,2017MNRAS.468.1671L,2020arXiv201008571S}.  While the mere existence of flattened structures is not seen as a major problem anymore \citep[e.g., ][]{2020arXiv200411585S}, the kinematic coherence of the satellite systems is a conundrum. Considering both phenomena, the spatial flattening and the kinematic coherence 
constitute the actual plane-of-satellites problem \citep{2018MPLA...3330004P}.

{Are flattened, co-rotating satellite systems a peculiarity of the Local Group or are they present around other galaxies as well?} Using the Sloan Digital Sky Survey (SDSS), \citet{2014Natur.511..563I} studied pairs of satellite galaxies on opposite sides of their host and found that their line-of-sight velocities are preferentially anti-correlated. {This is expected if they lie in co-rotating planes, suggesting that 
planes of satellites might be abundant in the Universe}, if the signal is indeed physical and not a statistical fluke (\citealt{2015MNRAS.453.3839P,2015MNRAS.449.2576C}, but see also \citealt{2015ApJ...805...67I}).

In the Local Volume ($D<10$ Mpc), there is evidence for flattened structures
around Cen\,A \citep{2015ApJ...802L..25T}, M\,83 \citep{MuellerTRGB2018}, and M\,101 \citep{2017A&A...602A.119M}. 
Regarding Cen\,A, \cite{2015ApJ...802L..25T} suggested the existence of two almost parallel {satellite planes}, but the subsequent discovery of new dwarf galaxies \citep{2016ApJ...823...19C,2017A&A...597A...7M} 
weakened the case {for} a strict separation {in} two planes \citep{2016A&A...595A.119M,MuellerTRGB2019}.
 Interestingly, the planes around Andromeda and Cen\,A are mainly aligned with the cosmic web, while the plane around the MW and the candidate planes around M\,83 and M\,101 are not \citep{2015MNRAS.452.1052L,2019MNRAS.490.3786L}. 

In \citet{2018Sci...359..534M} {we reported the discovery of a kinematic}
correlation among the satellites {of Cen\,A}{: 14 out of 16 galaxies seemingly co-rotate around the host}. {In 
cosmological simulations, satellite systems with a similar degree of kinematic coherence} {and flattening were found in only $\leq 0.5\%$ of Cen\,A analogs, indicating} a similar degree of conflict as in earlier studies {of the} {MW} {and Andromeda.} 
In  \citet{2020arXiv201104990M}, we have presented spectroscopy taken with the Multi Unit Spectroscopic Explorer (MUSE) mounted at the Very Large Telescope (VLT) of 12 additional dwarf galaxies around Cen\,A. Here we use them to test our previous assessment of a {co-rotating} plane-of-satellites around Cen\,A.

\section{The Cen\,A satellite system}\label{sec:dynamics}

\begin{table*}[ht]
\renewcommand{\arraystretch}{1.1}
\caption{Members of Cen\,A used in this study.  For KKs\,59 we adopted the same distance as Cen\,A because there is no accurate TRGB distance available. (a): galaxy name, (b): alternative PGC name, (c): right ascension in epoch J2000, (d): declination in epoch J2000, (e) galaxy distance, and reference for the distance measurement, (f): galaxy heliocentric velocity and  reference for the velocity measurement, (g): the technique of the velocity measurement, and (e): de Vaucouleurs morphological type according to the \om{Local Volume catalog \citep{2004AJ....127.2031K,2013AJ....145..101K}}.
}             % title of Table
\centering                          % used for centering table
\begin{tabular}{l l  c c l l l l}        % centered columns (4 columns)
\hline\hline                 % inserts double horizontal lines
Galaxy Name &  Alternative name &$\alpha_{2000}$ & $\delta_{2000}$ & $D$  & $v_h$ & Source & Type  \\    % table heading 
& & (degrees) & (degrees) &  (Mpc)   &(km\,s$^{-1}$)  \\    % table heading 
(a) & (b) & (c)& (d) & (e) & (f) & (g) & (e) \\    % table heading 
\hline      \\[-2mm]                  % inserts single horizontal line
ESO\,269-037 & PGC045916        &         195.8875&       $-$46.5842&   	3.15$\pm$0.09 (1)  &744$\pm2$ (2) & HI & dIrr\\
NGC\,4945  & PGC045279         &         196.3583&       $-$49.4711&       3.72$\pm$0.03 (3) &563$\pm3$ (4) & HI & Scd \\ %HIPASS 3.47
ESO\,269-058 & PGC045717        &         197.6333&       $-$46.9908&       3.75$\pm$0.02 (1) &400$\pm18$ (5) & HI &dIrr\\ %HIPASS
KK\,189    & PGC166158     &         198.1883&       $-$41.8320 &       4.21$\pm$0.17 (6) &753$\pm4$  (7) & stars & dSph \\
ESO\,269-066  & PGC045916      &         198.2875&       $-$44.8900&       3.75$\pm$0.03 (1) &784$\pm31$ (8)& stars& dSph\\
NGC\,5011C & PGC045917         &         198.2958&       $-$43.2656&       3.73$\pm$0.03 (1) &647$\pm96$ (9)& stars & Tr\\
KKs\,54 &   PGC2815821      &         200.3829&       $-$31.8864 &   3.75$\pm$0.10      (10) &621$\pm11$  (7)& stars & Tr \\
KK\,196 & PGC046663            &         200.4458&       $-$45.0633&       3.96$\pm$0.11 (1) &741$\pm15$ (11)& stars & dIrr\\
NGC\,5102 & PGC046674          &         200.4875&       $-$36.6297&       3.74$\pm$0.39 (3)&464$\pm18$ (12) & HI & Sa\\ %HIPASS 3.66
KK\,197  & PGC046680      &         200.5086&       $-$42.5359 &   3.84$\pm$0.04 (1)  &643$\pm3$  (7) & stars& dSph \\
KKs\,55 & PGC2815822       &        200.5500 &       $-$42.7311 &   3.85$\pm$0.07(1)  &530$\pm14$  (7) & stars & Sph \\
dw1322-39 &        &        200.6336 &       $-$39.9060 &   2.95$\pm$0.05  (10) &656$\pm10$  (7) & stars &  dIrr\\
dw1323-40b &        &        200.9809&       $-$40.8361 &   3.91$\pm$0.61  (10) &497$\pm12$  (7) & stars & dSph \\
dw1323-40a &       &       201.2233&       $-$40.7612 &   3.73$\pm$0.15   (10)&450$\pm14$  (7) & stars & dSph\\
\it Cen\,A  & PGC046957         &    \it 201.3667&\it    $-$43.0167&\it    3.68$\pm$0.05  (1) & \it 556$\pm$10 (4) & HI & S0$^{0}$\\
KK\,203 &   PGC166167    &       201.8681&       $-$45.3524 &    3.78$\pm$0.25 	  (3) &306$\pm10$  (7)& stars & Tr \\
ESO\,324-024 & PGC047171       &         201.9042&       $-$41.4806&       3.78$\pm$0.09 (1)&514$\pm18$ (12) & HI & Sdm \\ %HIPASS
NGC\,5206 & PGC047762          &         203.4292&       $-$48.1511&       3.21$\pm$0.01 (1) &583$\pm6$ (13) & stars & S0$^{-}$\\
NGC\,5237 & PGC048139          &         204.4083&       $-$42.8475&       3.33$\pm$0.02 (1)  &361$\pm4$ (4) & HI & BCD \\ %HIPASS
NGC\,5253 & PGC048334           &         204.9792&       $-$31.6400&       3.55$\pm$0.03 (3) &407$\pm3$ (4) & HI & Sdm \\ %HIPASS 3.44
dw1341-43 &        &       205.4032 &       $-$43.8553 &    3.53$\pm$0.04 	(10)  &636$\pm14$  (7)& stars& dSph \\
KKs\,57 & PGC2815823      &       205.4079 &       $-$42.5797 &     3.84$\pm$0.47	(1)  &511$\pm17$  (7)& stars & Sph\\
KK\,211 & PGC048515             &         205.5208&       $-$45.2050&       3.68$\pm$0.14 (1) &600$\pm31$ (14) & stars & Sph \\
dw1342-43  &     &      205.6837 &       $-$43.2548 &      2.90$\pm$0.14	  (10) &510$\pm8$  (7)& stars & Tr \\
ESO\,325-011  & PGC048738      &         206.2500&       $-$41.8589&       3.40$\pm$0.05 (1) &544$\pm1$ (15)& HI & dIrr\\ %HIPASS
KKs\,58  & PGC2815824     &    206.5031 &       $-$36.3289 &      3.36$\pm$0.10	  (10) &477$\pm5$   (7)& stars&dSph \\
KK\,221 & PGC166179            &         207.1917&       $-$46.9974&       3.82$\pm$0.07 (1) &507$\pm13$ (14) & stars& dIrr\\
ESO\,383-087  & PGC049050       &         207.3250&       $-$36.0614&       3.19$\pm$0.03 (1) &326$\pm2$ (4) & HI& Sdm \\ %HIPASS
ESO174-001/KKs\,57 &   PGC048937               &         206.9920&       $-$53.3476&       3.68* &686$\pm1$ (15) & HI& dIrr\\
\hline
\end{tabular}
\tablefoot{The references are: (1): \citet{2013AJ....145..101K},  (2): \citet{2007AJ....133..261B}, (3): \citet{2015ApJ...802L..25T}, (4): \citet{2004AJ....128...16K}, (5): \cite{1999ApJ...524..612B}, (6): \cite{2008ApJ...676..184T}, (7): \citet{2020arXiv201104990M}, (8): \cite{2000AJ....119..166J}, (9): \cite{2007AJ....133.1756S}, (10): \cite{MuellerTRGB2019}, (11): \cite{2000AJ....119..593J}, (12): \cite{2005MNRAS.361...34D}, (13): \cite{1993AJ....105.1411P}, (14): \cite{2008ApJ...674..909P}, and (15): \cite{2012MNRAS.420.2924K}. }
\label{tab:sample}
\end{table*}

\begin{figure*}[ht]
    \centering
    \includegraphics[width=\linewidth]{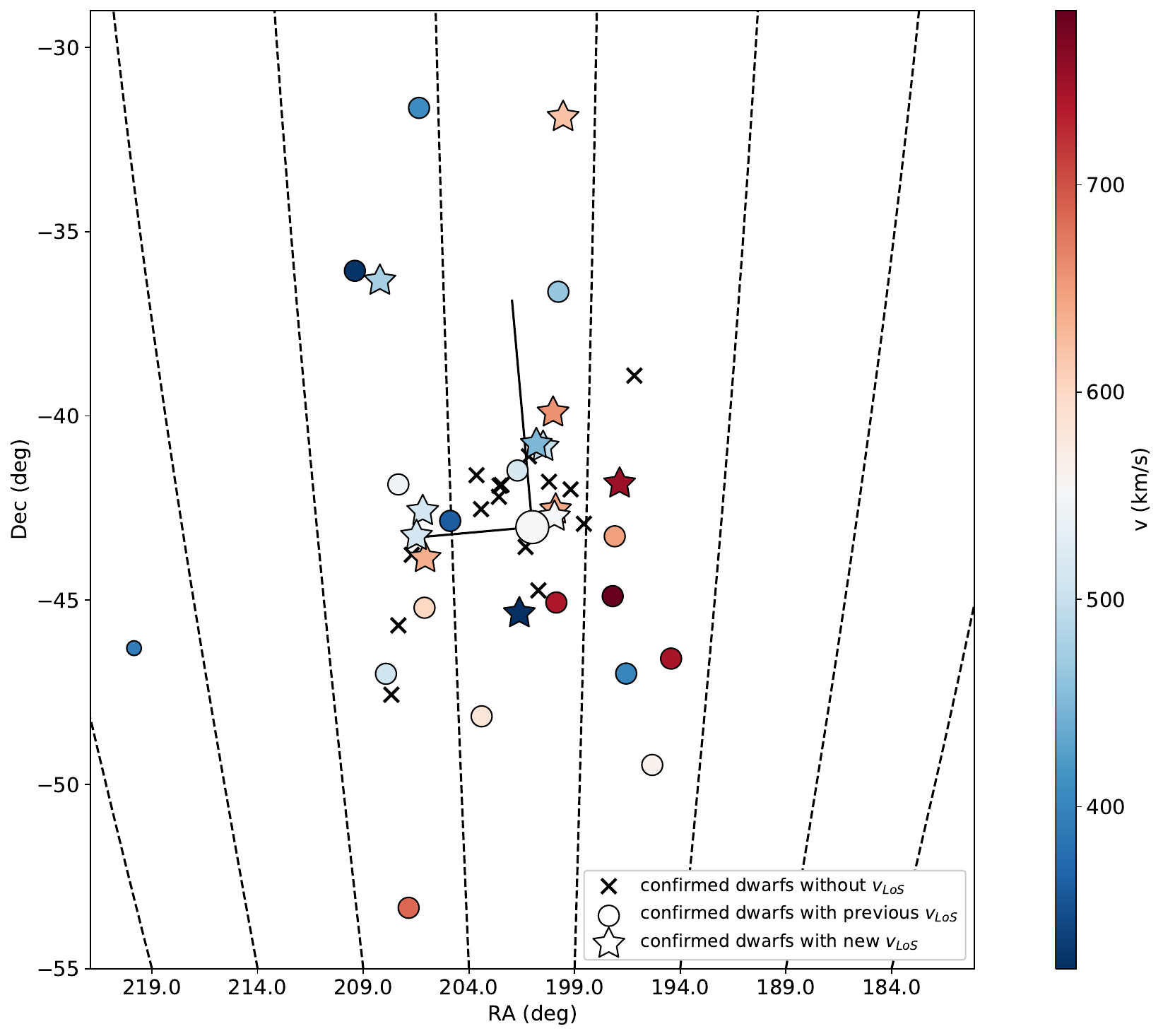}
    \caption{The on-sky distribution of the Cen\,A {satellite} system within 800\,kpc. 
    The {circles} correspond to the dwarf galaxies studied in \citet{2018Sci...359..534M}, the stars to the newly observed dwarfs \citep{2020arXiv201104990M}. 
    The colors indicate whether the galaxies are red-, or blue-shifted with respect to the systemic velocity of Cen\,A (shown with large open circle). The crosses are dwarf galaxies with known distances but without velocity information, and the small blue dot to the left is a dwarf galaxy not belonging to the plane-of-satellites.
    The \om{black lines} centered on Cen\, A corresponds to the major and minor axes of the satellite distribution.
    }
     \label{fig:field}
\end{figure*}

\begin{figure}[ht]
    \centering
    \includegraphics[width=\linewidth]{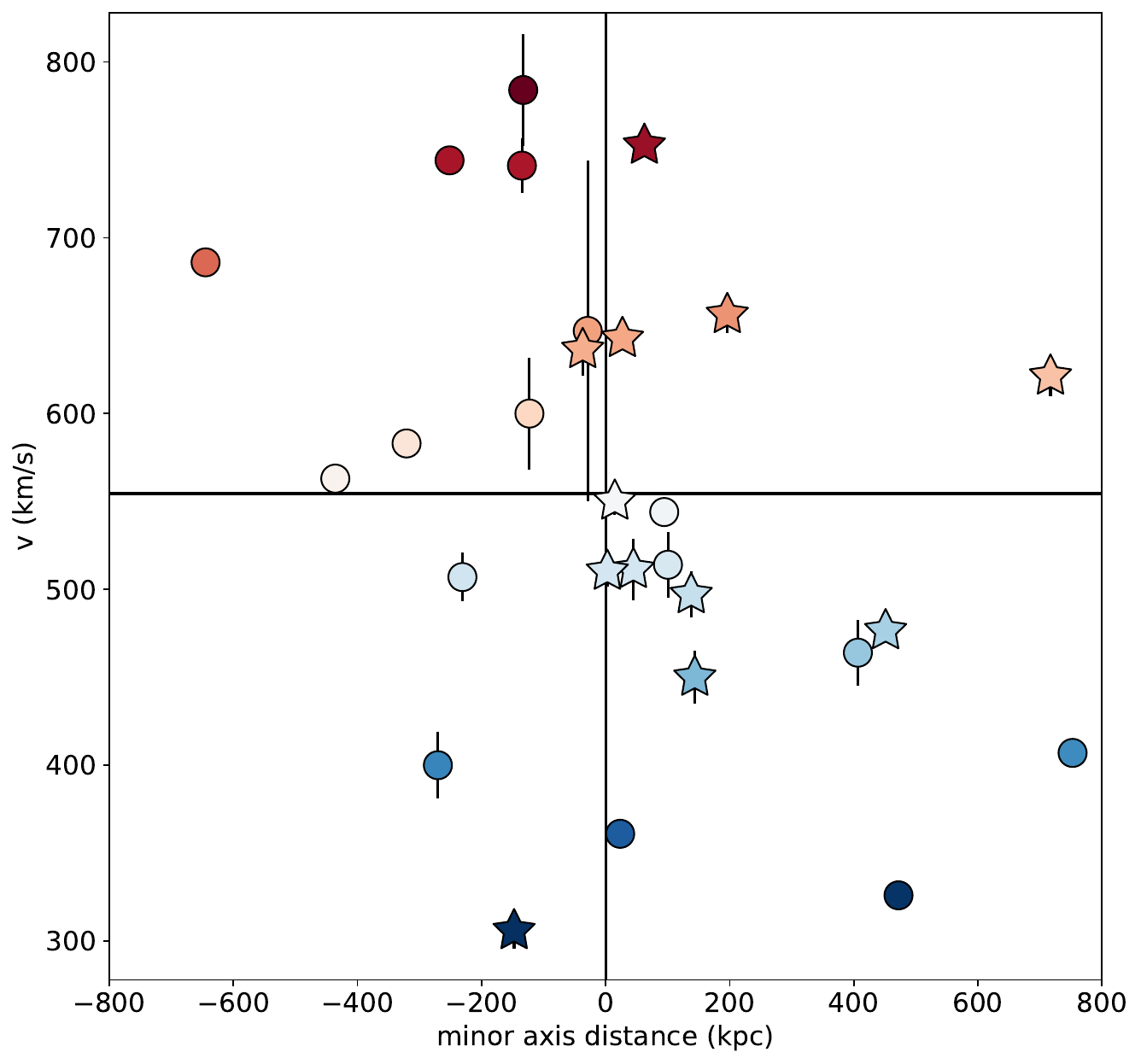}
    \caption{Position-velocity diagram for the dwarf galaxy satellite system of Cen\,A. \om{The x-axis  represents the distance from the minor axis (i.e. along the major axis) derived from the satellite distribution.} The filled circles show the dwarf galaxies used in \citet{2018Sci...359..534M}, and the stars illustrate the new data. \om{The uncertainties are always plotted, but often smaller than the dots.} Approaching and receding galaxies with respect to the mean of the group are shown in blue and red, respectively, as in Fig\,\ref{fig:field}.  }
    \label{fig:pv}
\end{figure}

Cen\,A -- the central galaxy of the {Centaurus} group -- is one of the best studied nearby galaxies beyond the Local Group.
{To date, 27 dwarf satellites of Cen\,A have both accurate distances and velocity measurements. The line-of-sight velocities were measured using two different techniques: emission lines from atomic and/or ionized gas for star forming galaxies (e.g. \citealt{2004AJ....128...16K}) or absorption line spectroscopy of the stars (e.g. \citealt{2008ApJ...674..909P}, \citealt{2020arXiv201104990M}).
The distances {come} from tip of the red giant branch (TRGB) measurements based on  HST programs 
    \citep{2007AJ....133..504K} {and} our VLT program \citep{MuellerTRGB2019}. An additional dwarf galaxy (ESO174-001) has a velocity measurement, but no accurate distance estimate. As we are mainly {focusing on} the motion of satellites, we include this dwarf in our sample,
    {for a grand total of 28 satellites}. All galaxies used in the analysis are presented in Table\,\ref{tab:sample}.
    
There are 13 more dwarfs around Cen\,A which have TRGB measurements  (e.g., \citealt{2019ApJ...872...80C}), but lack the velocities necessary to study the kinematics of the satellite system. {More than 30 additional candidates} \citep{2015A&A...583A..79M,2017A&A...597A...7M,2018ApJ...867L..15T} await {membership} confirmation.
    }

Figure \ref{fig:field} shows the on-sky distribution {and line-of-sight velocities} of the {28} satellites 
with respect to Cen\,A. {If the co-rotating satellite system suggested by \citet{2018Sci...359..534M} is a real physical structure}, then the dwarfs to the north of Cen\,A should be blue-shifted, {while} the dwarfs to the south should be red-shifted. This {is}
the case for 21 out of 28 of the dwarf satellites.

Figure\,\ref{fig:pv} presents the position-velocity diagram for the dwarf galaxies around Cen\,A. {For convenience, we assign positive and negative separations from Cen\,A to the Northern and Southern satellites, respectively, adopting a separating line with a position angle $PA=95^{\circ}$ and the mean velocity of the group of $v=555$\,km\,s$^{-1}$.} This position angle corresponds to the geometric minor axis of \om{the 28 satellite galaxies studied in this work.}
If the satellite system {has significant rotational support} with all the members {having near-circular orbits and} sharing the same orbital {motion,}
we would expect them to populate two opposing quadrants in the position-velocity diagram. On the other hand, a pressure-supported system would fill all four quadrants equally. {Figure \ref{fig:pv}} clearly favors the former {case}: {out of} 28 galaxies,
21 lie in two opposing quadrants.

{If we assume} that this signal arises by chance and calculate {its} probability via the binomial coefficient, 
{the probability of} finding exactly 21 out of 28 galaxies in opposing quadrants is $P(X=21$ $|$ $28)=0.88$\%, while the one of finding 21 or more out of 28 is $P(X\geq21$ $|$ $28)=1.26$\%.
For comparison, \citet{2018Sci...359..534M} found 14 out of 16 satellites in opposing quadrants yielding a probability of $P(X=14 $ $|$ $16)=0.36$\% or $P(X\geq14 $ $|$ $16)=0.42$\%. {Thus, the statistical significance of a kinematic signal has nearly stayed the same after almost doubling the sample size.} Therefore, it is likely that a rotational component is present in the satellite system -- even though 5 of the new data points are not following this pattern.
{Two of the latter lie in the inner region of the system, where the line-of-sight velocities are less likely to indicate the orbital direction if the satellites follow non-circular orbits.}

\section{Comparison to cosmological expectations}\label{sec:cosmo}
The high degree of kinematic correlation among the satellite galaxies of Cen\,A was {shown to be}
rare in cosmological simulations \citep{2018Sci...359..534M}. In the hydrodynamical Illustris simulation \citep{2014MNRAS.444.1518V}, only 0.5 per cent of mock-observed satellite systems around hosts of similar virial mass as Cen\,A were both as flattened and {kinematically correlated}. For the dark-matter-only Millennium-II simulation \citep{2009MNRAS.398.1150B}, the frequency of analogs to the Cen\,A satellite system was even lower: {only} 0.1 per cent. 
Here we update these comparisons adding our new data to determine whether the tension with cosmological expectations remains or is alleviated given the more {comprehensive} observational {picture}.

We base our comparison on the IllustrisTNG project, specifically the TNG100-1 run \citep[e.g.,][]{2018MNRAS.475..676S,2018MNRAS.475..648P}.
%\citep{2018MNRAS.477.1206N, 2018MNRAS.475..676S,2018MNRAS.480.5113M, 2018MNRAS.475..624N,2018MNRAS.475..648P}.
IllustrisTNG expands the original Illustris simulations by refining the implemented physics. 
The adopted cosmological parameters are consistent with a Planck cosmology \citep{2016A&A...594A..13P}. The simulations box size ($75\,\mathrm{Mpc}/h$), and resolution (dark matter particle mass $m_\mathrm{DM} = 7.5 \times 10^6\,\mathrm{M}_\odot$\ for the hydrodynamical run) provides a good compromise between the number of {hosts and the} number of resolved satellites.
We use the publicly available redshift zero galaxy catalogs \citep{2019ComAC...6....2N}. By using both the hydrodynamical and the equivalent dark-matter-only (DMO) run we can directly determine whether baryonic physics (as implemented in the simulation) affects the planes-of-satellites issue.

\subsection{Simulated Cen\,A analogs and mock satellite systems}

We select Cen\,A analog host galaxies by a mass and isolation criterion. The simulated host halos are required to have a virial mass $M_{200}$\ between $4 \times 10^{12}$\ and $12 \times 10^{12}\,\mathrm{M}_\odot$. This 
is identical to the mass range of Cen\,A analogs in \citet{2018Sci...359..534M}. We require each host to be isolated by rejecting any potential hosts which contain a second halo with virial mass $M_{200} \geq 0.5 \times 10^{12} M_\sun$\ within 
1.2\,Mpc radius, {motivated by the distance of Cen\,A to M\,83, the second major galaxy in the Centaurus group.}
For each host we identify all (subhalo) galaxies within 
800\,kpc as possible satellites{, comparable to the distance range of observed satellites (see Fig.~\ref{fig:pv})}. Note that we only select by this volume and do not require the possible satellites to be bound to their host, since no such selection was made  on the observational data either.
Finally, to exclude cases of possible ongoing major mergers, we reject all hosts which have a satellite whose stellar mass exceeds $1/4$\ of the host galaxy stellar mass, as these would not resemble the observed Cen\,A system.
For the DMO run, the latter criterion is applied to the virial masses, though a smaller number of hosts is rejected in this case since dark matter mass is more easily stripped and thus reduced faster after infall. These criteria result in 180 Cen\,A analogs from the hydrodynamical TNG100 run and 307 from the the DMO run.

In the DMO run the satellites of a given host are ranked by their dark matter mass. In the hydrodynamical run the satellites are ranked by their stellar mass first, and then by their dark matter mass for those subhalos that do not contain stellar particles. We allow the inclusion of such dark satellites to ensure a large sample of satellite systems with a sufficient number of satellites to compare to the observed situation, as was also done for {MW} analogs in \citet{2020MNRAS.491.3042P}.
It is safe to assume that the presence or absence of stars in subhalos of dwarf-galaxy scale does not strongly affect the subhalo position and motion, given the dynamical dominance of the dark matter component. 
While this ensures a meaningful test  of cosmological predictions, it could be improved upon in the future by comparisons with dedicated, higher-resolution zoom simulations of Cen\,A analogs.

\begin{figure*}[ht]
    \centering
    \includegraphics[width=0.45\linewidth]{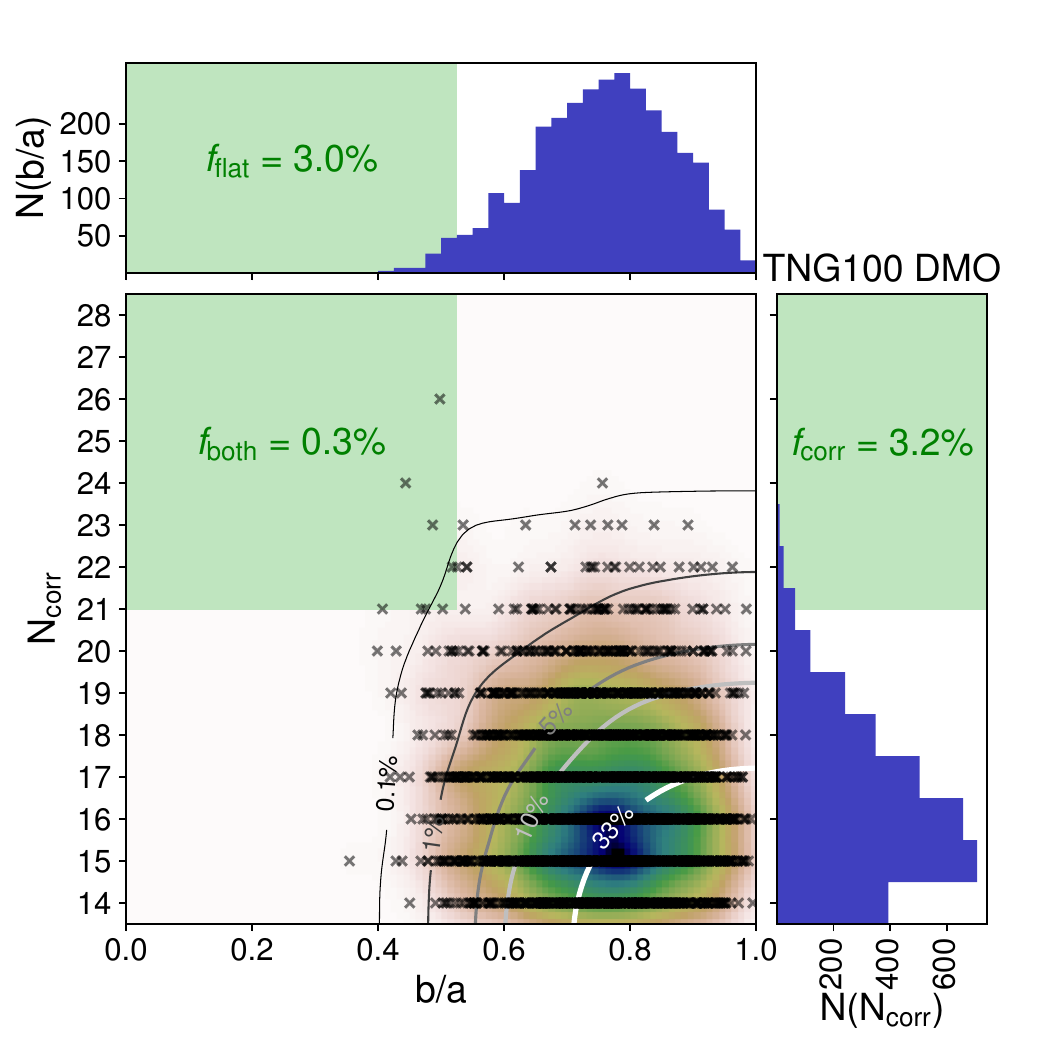}
    \includegraphics[width=0.45\linewidth]{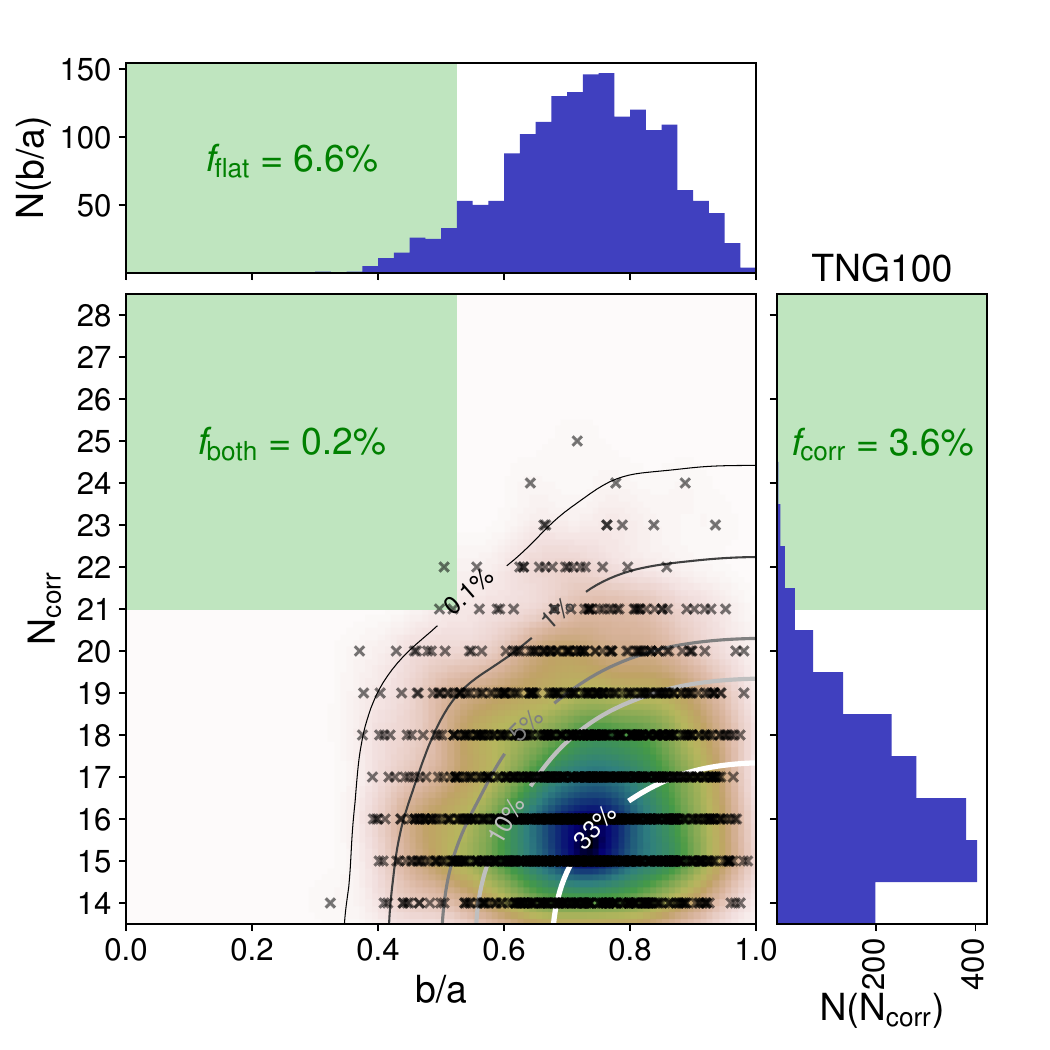}
    \caption{{Mock-observed satellite systems around Cen\,A analogs from the Illustris TNG-100 simulation, considering both its dark-matter-only run (left) and hydrodynamical run (right).} 
    The vertical axis plots the number of kinematically correlated satellites $N_{\rm corr}$, with $N_{\rm corr}=21$\ for the observed Cen\,A system. The horizontal axis plots the on-sky axis-ratio flattening $b/a$. The histograms show the number of realizations with a given axis ratio $N(b/a)$\ and a given number of correlated velocities $N(N_{\rm corr})$, respectively.
    The color maps indicate the density of simulated systems, while the contours indicate what fraction of simulated systems are more extreme than the parameter combination.   The green shaded regions indicate parameter combinations that are as extreme or more as the observed Cen\,A system, while $f_{\rm flat}$, $f_{\rm corr}$, and $f_{\rm both}$\ report the fraction of mock systems that are at least as flattened as the observed system, at least as kinematically correlated, or both simultaneously.
    }
    \label{fig:TNG}
\end{figure*}

We mock-observe each satellite system by placing the host at a distance of 3.68\,Mpc, projecting the satellites on the sky in angular coordinates relative to the host and calculating the line-of-sight velocity component of each satellite. Satellites are then selected in order from the ranked list requiring them to be between $1^\circ$\ and $12^\circ$\ from the host. For a comparison to the extended data set presented in this study we require 28 satellites to be selected. 
For each of the 180 hosts identified in the hydrodynamic TNG100 simulation, this process is repeated for ten different random viewing directions. We obtain 1763 mock-observed systems, because for 37 realizations {there were less than}
28 simulated satellites 
within the mock-observed volume around their host. For the DMO {run,}
we have 307 hosts and 3070 mock-observed realizations.

We checked that the selected mock satellite systems have a radial {root-mean-square} (rms) distribution in 
angular coordinates {on the sky} that is consistent with that of the observed Cen\,A system. For the 28 observed satellites,
we find a rms radial distance from the position of Cen\,A of $R_\mathrm{rms} = 5\fdg46 \pm 0\fdg67$ with the uncertainty determined by bootstrapping. From the hydrodynamical run, the average and standard deviation of the simulated systems with 28 satellites is $\langle R_\mathrm{rms} \rangle = 5\fdg69 \pm 0\fdg64$. The DMO run tends to result in slightly more radially extended satellite systems, with an average and standard deviation of $\langle R_\mathrm{rms}\rangle = 6\fdg33 \pm 0\fdg66$. \om{This has a couple of reasons. The combined effect of a steeper inner density profile due to a disk, less mass loss due to tides, and a larger dynamical friction means that halos have more concentrated satellite distributions in hydrodynamical than DMO simulations \citep{2010MNRAS.401.1889L}.}

\subsection{Frequency of Cen\,A analogs in simulations}

As in \citet{2018Sci...359..534M}, we decide to be conservative and consider only the on-sky distribution of the simulated satellite systems and their line-of-sight velocity components.
For each mock-observed system, we find the major axis of the on-sky satellite distribution and measure the rms flattening perpendicular to this direction (the minor to major axis ratio $b/a$). We also count the number $N_\mathrm{corr}$\ of satellite galaxies with coherent line-of-sight velocities along the major axis {(i.e., consistently redshifted or blueshifted with respect to the central halo)} by dividing the on-sky satellite distribution along {the minor axis.}
For the observed Cen\,A system, this yields an on-sky flattening of $b/a = 0.52$\ and $N_\mathrm{corr} = 21$.
A slightly larger $N_\mathrm{corr}$\ number might be found for a different orientation of the dividing line, but to ensure an unbiased comparisons to the simulations we opt to define the orientation using only spatial information. 

The results of our comparison for the full sample of 28 satellites is shown in Fig.~\ref{fig:TNG}. \om{Taking the frequency that both the kinematic coherence and flattening is as extreme or more as the observed Cen\,A satellite population (measured as  $f_\mathrm{both}$),}
we find that only $f_\mathrm{both} = 0.3$\% of the DMO mock systems (8 out of 3070 realizations), and $f_\mathrm{both} = 0.2$\% of the hydrodynamic mock systems (3 out of 1763 realizations) are as extreme or more as the  Cen\,A system.
{Thus, the inclusion of baryon physics does not alleviate the plane-of-satellites problem for Cen\,A.}
Even though the fraction of correlated satellites has dropped compared to the earlier study the significance or rarity of this occurring in cosmological simulations remains unchanged, since we are probing a larger fraction of the total satellite population. 

The fraction $f_\mathrm{corr}$ of simulated systems which are at least as kinematically correlated  as observed is consistent between the DMO and hydrodynamical runs (3.2 versus 3.6\%). However, the DMO run has a smaller frequency of similarly flattened satellite systems ($f_\mathrm{flat} = 3.0$\%) than the hydrodynamic run ($f_\mathrm{flat} = 6.6$\%).
We suspect this is related to the slightly less compact radial distribution in the DMO run, as it has been found that more compact satellite systems tend to result in more flattened distributions \citep{2019ApJ...875..105P}.

\om{For the previous analysis, we have taken the 28 brightest (or most massive in the DMO simulation) subhalos to compare our observations to. Does this ranked selection of subhalos bias the results? To assess this, we have repeated our analysis on the TNG-100 simulation, considering the hydrodynamical run. Instead of selecting the 28 brightest subhalos, we selected the 40, 45, and 50 brightest subhalos and draw a random sample of 28 subhalos out of those.  Finding 21 out of 28 coherently moving and flattened satellites as observed around Cen\,A occurs in 0.3\%, 0.4\%, and 0.1\% of the simulated Cen\,A analogs, respectively. In other words, the frequency doesn't change whether we pick the 28 brightest subhalos or draw randomly from a larger sample of subhalos. We drew the same conclusion in \citet{2018Sci...359..534M}.
}

\subsection{Properties of identified analogs}

\begin{figure*}[ht]
    \centering
    \includegraphics[width=0.33\linewidth]{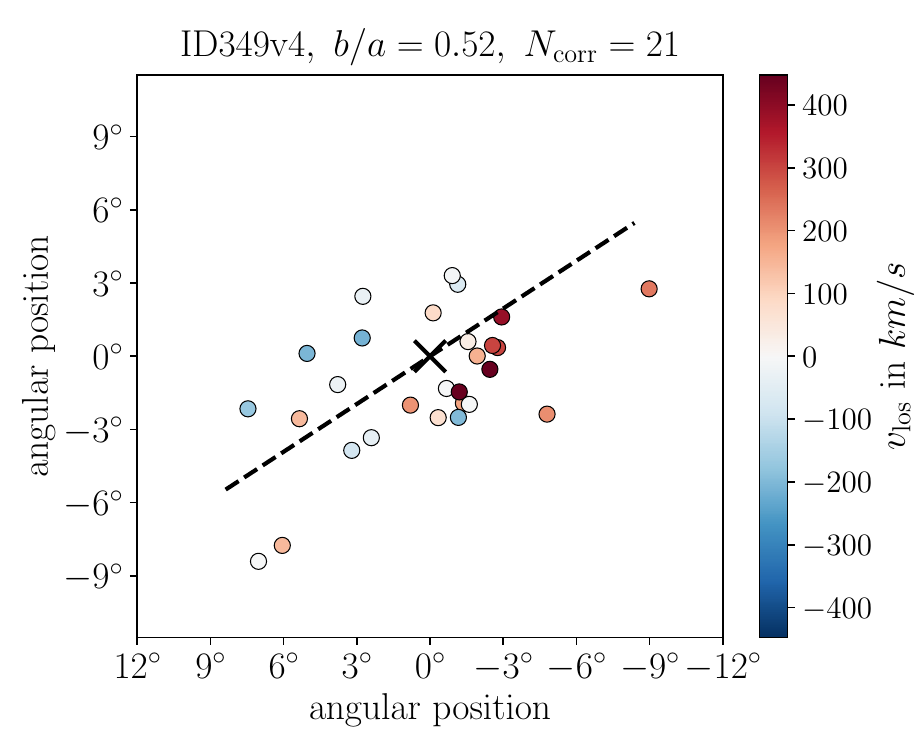}
    \includegraphics[width=0.33\linewidth]{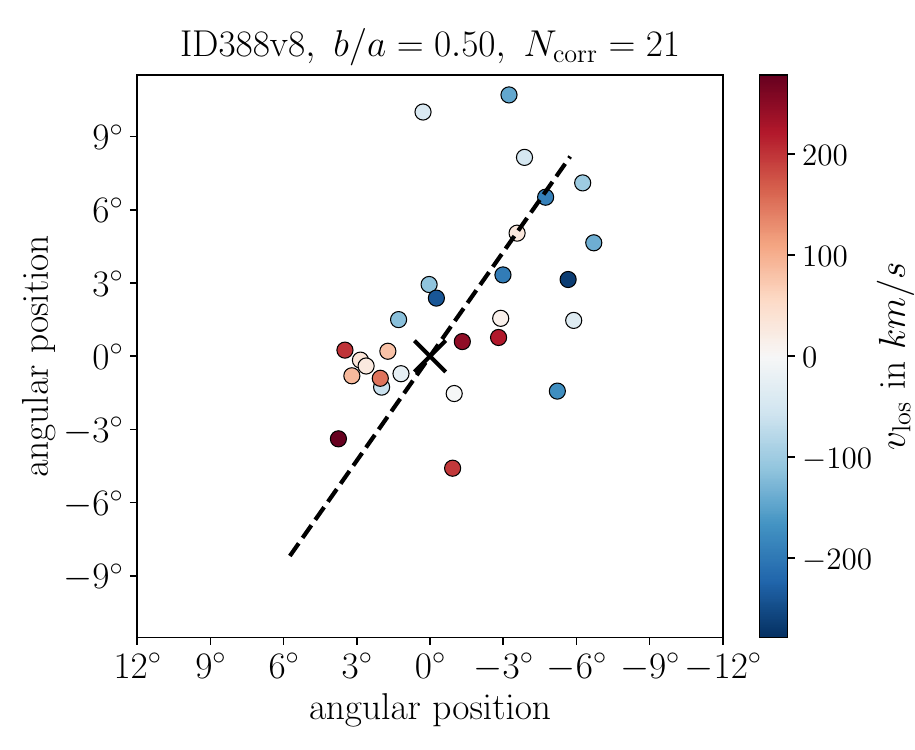}
    \includegraphics[width=0.33\linewidth]{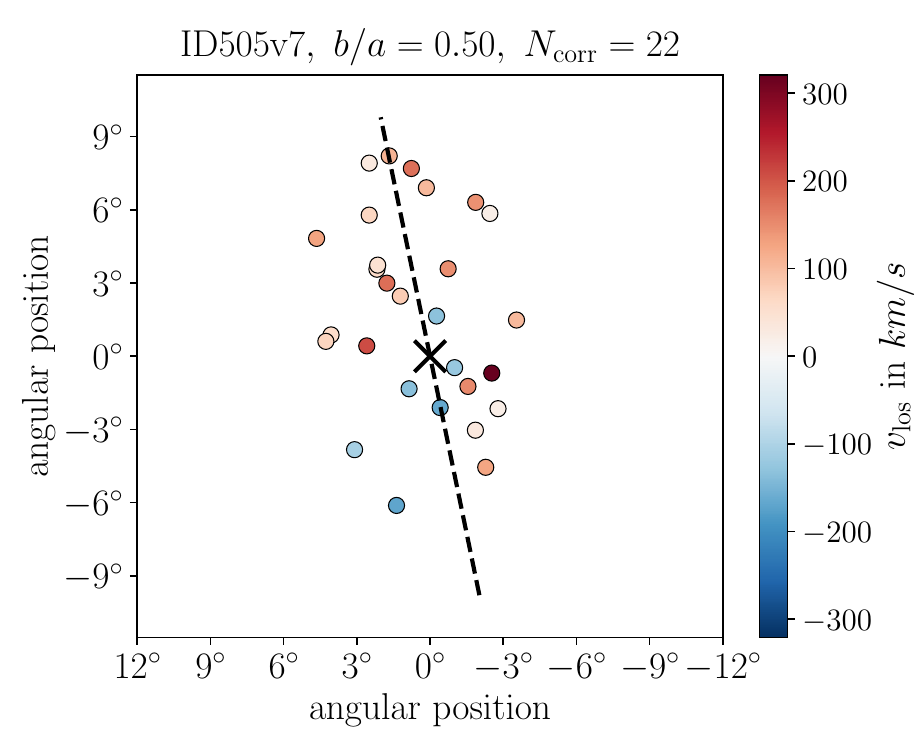}
    \caption{Mock-observed on-sky satellite distributions of the three simulated systems that have satellite plane parameters as extreme as observed. The satellite galaxies (circles) are color coded by their line-of-sight velocity relative to the host (black cross). The dashed lines indicate the major axis of the projected distribution of satellites. The top labels give the minor-to-major axis flattening $b/a$\ of the on-sky distribution for the simulated systems, as well as the number of correlated velocities $N_\mathrm{corr}$.
    \label{fig:simsystems}}
\end{figure*}

Fig. \ref{fig:simsystems} shows the three mock satellite systems in the hydrodynamical Illustris TNG simulation that are at least as flattened ($b/a\leq 0.52$) and have at least as many kinematically correlated satellites ($N_\mathrm{corr}\geq 21$) as Cen\,A. For each of these three cases, only one out of 10 random view directions results in a match. The other nine random view directions do not simultaneously reproduce the observed spatial flattening and kinematic correlation. 
{This {suggests}
that planes of satellites in $\Lambda$CDM simulations are of transient nature {due to chance projections \citep[see also][]{2016MNRAS.460.4348B}. They} do not form co-rotating structures, otherwise we would expect to mock-observe {a similar} kinematic coherence from different directions.} Therefore, the probability of having chance projections in the three nearest and best-studied satellite systems (Milky Way, Andromeda, and Cen\,A) appears arbitrarily small.

The central halos of the three \om{simulated systems resembling the Cen\,A satellite population} have varied merger histories. Two of them show quiescent growth over the last 10 Gyr, while one experienced a major merger 4 Gyr ago. The high coherence in the latter's satellites is unlikely to result from the merger event, contrary to what was proposed by \citet{Smith2016}. Most of a major merger influence would have been washed out by the stripping of participating satellites and the accretion of new satellites since (Kanehisa et al. in prep). None of these three analogs match Cen A's reported major merger 2 Gyr ago \citep{Wang2020}.

\section{Summary and conclusions}\label{sec:summ}

The phase-space distribution of dwarf satellite galaxies provides a key testbed
for cosmological predictions on small scales. For  the Cen\,A system it has been suggested based on 16 line-of-sight velocities \citep{2018Sci...359..534M} that most dwarf galaxies are aligned in a co-rotating structure similar to what has been
found for the MW and the Andromeda galaxy. {We revisited this issue by adding 12 more satellites of Cen\,A, for which we acquired line-of-sight velocities using MUSE spectroscopic observations \citep{2020arXiv201104990M}. This increases the sample size from 16 to 28 satellites and enhances the representation of gas-poor dwarf spheroidal galaxies. %Our results can be summarized as follows:
}

{We find that 21 out of 28 satellites show a coherent motion.} This implies that the {co-moving satellite system inferred around Cen\,A was not a  statistical fluke} due to the small number of tracers, but is a real phenomenon.

{In the Illustris-TNG simulation, satellite systems that are at least as flattened {\it and} kinematically correlated as Cen\,A occur with a} {frequency} {of only 0.2$\%$ in the hydrodynamical and 0.3$\%$ in the DMO run. This indicates that the plane-of-satellites problem is independent of baryon physics.} {The simulated satellite systems that are consistent with the observed Cen\,A properties are not stable, co-rotating structures: their kinematic coherence is arising by chance projection.}

In summary, we find that the additional kinematic data obtained for 12 more Cen\,A satellites \citep{2020arXiv201104990M} does not alleviate the tension with $\Lambda$CDM expectations. The new velocities further support the findings of \citet{2018Sci...359..534M}. The satellite {system of Cen\,A} remains a challenge for $\Lambda$CDM.

\begin{acknowledgements} 
{We thank the referee for the constructive report, which helped to clarify and improve the manuscript.}
O.M. is grateful to the Swiss National Science Foundation for financial support.
M.S.P. and O.M. thank the DAAD for PPP grant 57512596 funded by the BMBF, and the Partenariat Hubert Curien (PHC) for PROCOPE project 44677UE. M.S.P. thanks the Klaus Tschira Stiftung gGmbH and German Scholars Organization e.V. for support via a Klaus Tschira Boost Fund. H.J. acknowledges financial support from the Australian Research Council through the Discovery Project DP150100862.
\end{acknowledgements}

\bibliographystyle{aa}
\bibliography{bibliographie}

\begin{thebibliography}{68}
\expandafter\ifx\csname natexlab\endcsname\relax\def\natexlab#1{#1}\fi

\bibitem[{{Banks} {et~al.}(1999){Banks}, {Disney}, {Knezek}, {Jerjen},
  {Barnes}, {Bhatal}, {de Blok}, {Boyce}, {Ekers}, {Freeman}, {Gibson},
  {Henning}, {Kilborn}, {Koribalski}, {Kraan-Korteweg}, {Malin}, {Minchin},
  {Mould}, {Oosterloo}, {Price}, {Putman}, {Ryder}, {Sadler}, {Staveley-Smith},
  {Stewart}, {Stootman}, {Vaile}, {Webster}, \& {Wright}}]{1999ApJ...524..612B}
{Banks}, G.~D., {Disney}, M.~J., {Knezek}, P.~M., {et~al.} 1999, \apj, 524, 612

\bibitem[{{Bouchard} {et~al.}(2007){Bouchard}, {Jerjen}, {Da Costa}, \&
  {Ott}}]{2007AJ....133..261B}
{Bouchard}, A., {Jerjen}, H., {Da Costa}, G.~S., \& {Ott}, J. 2007, \aj, 133,
  261

\bibitem[{{Boylan-Kolchin} {et~al.}(2009){Boylan-Kolchin}, {Springel}, {White},
  {Jenkins}, \& {Lemson}}]{2009MNRAS.398.1150B}
{Boylan-Kolchin}, M., {Springel}, V., {White}, S.~D.~M., {Jenkins}, A., \&
  {Lemson}, G. 2009, \mnras, 398, 1150

\bibitem[{{Buck} {et~al.}(2016){Buck}, {Dutton}, \&
  {Macci{\`o}}}]{2016MNRAS.460.4348B}
{Buck}, T., {Dutton}, A.~A., \& {Macci{\`o}}, A.~V. 2016, \mnras, 460, 4348

\bibitem[{{Cautun} {et~al.}(2015){Cautun}, {Wang}, {Frenk}, \&
  {Sawala}}]{2015MNRAS.449.2576C}
{Cautun}, M., {Wang}, W., {Frenk}, C.~S., \& {Sawala}, T. 2015, \mnras, 449,
  2576

\bibitem[{{Crnojevi{\'c}} {et~al.}(2019){Crnojevi{\'c}}, {Sand}, {Bennet},
  {Pasetto}, {Spekkens}, {Caldwell}, {Guhathakurta}, {McLeod}, {Seth}, {Simon},
  {Strader}, \& {Toloba}}]{2019ApJ...872...80C}
{Crnojevi{\'c}}, D., {Sand}, D.~J., {Bennet}, P., {et~al.} 2019, \apj, 872, 80

\bibitem[{{Crnojevi{\'c}} {et~al.}(2016){Crnojevi{\'c}}, {Sand}, {Spekkens},
  {Caldwell}, {Guhathakurta}, {McLeod}, {Seth}, {Simon}, {Strader}, \&
  {Toloba}}]{2016ApJ...823...19C}
{Crnojevi{\'c}}, D., {Sand}, D.~J., {Spekkens}, K., {et~al.} 2016, \apj, 823,
  19

\bibitem[{{Dinescu} {et~al.}(2004){Dinescu}, {Keeney}, {Majewski}, \&
  {Girard}}]{2004AJ....128..687D}
{Dinescu}, D.~I., {Keeney}, B.~A., {Majewski}, S.~R., \& {Girard}, T.~M. 2004,
  \aj, 128, 687

\bibitem[{{Doyle} {et~al.}(2005){Doyle}, {Drinkwater}, {Rohde}, {Pimbblet},
  {Read}, {Meyer}, {Zwaan}, {Ryan-Weber}, {Stevens}, {Koribalski}, {Webster},
  {Staveley-Smith}, {Barnes}, {Howlett}, {Kilborn}, {Waugh}, {Pierce},
  {Bhathal}, {de Blok}, {Disney}, {Ekers}, {Freeman}, {Garcia}, {Gibson},
  {Harnett}, {Henning}, {Jerjen}, {Kesteven}, {Knezek}, {Mader}, {Marquarding},
  {Minchin}, {O'Brien}, {Oosterloo}, {Price}, {Putman}, {Ryder}, {Sadler},
  {Stewart}, {Stootman}, \& {Wright}}]{2005MNRAS.361...34D}
{Doyle}, M.~T., {Drinkwater}, M.~J., {Rohde}, D.~J., {et~al.} 2005, \mnras,
  361, 34

\bibitem[{{Ibata} {et~al.}(2014{\natexlab{a}}){Ibata}, {Ibata}, {Famaey}, \&
  {Lewis}}]{2014Natur.511..563I}
{Ibata}, N.~G., {Ibata}, R.~A., {Famaey}, B., \& {Lewis}, G.~F.
  2014{\natexlab{a}}, \nat, 511, 563

\bibitem[{{Ibata} {et~al.}(2015){Ibata}, {Famaey}, {Lewis}, {Ibata}, \&
  {Martin}}]{2015ApJ...805...67I}
{Ibata}, R.~A., {Famaey}, B., {Lewis}, G.~F., {Ibata}, N.~G., \& {Martin}, N.
  2015, \apj, 805, 67

\bibitem[{{Ibata} {et~al.}(2014{\natexlab{b}}){Ibata}, {Ibata}, {Lewis},
  {Martin}, {Conn}, {Elahi}, {Arias}, \& {Fernando}}]{2014ApJ...784L...6I}
{Ibata}, R.~A., {Ibata}, N.~G., {Lewis}, G.~F., {et~al.} 2014{\natexlab{b}},
  \apjl, 784, L6

\bibitem[{{Ibata} {et~al.}(2013){Ibata}, {Lewis}, {Conn}, {Irwin},
  {McConnachie}, {Chapman}, {Collins}, {Fardal}, {Ferguson}, {Ibata}, {Mackey},
  {Martin}, {Navarro}, {Rich}, {Valls-Gabaud}, \&
  {Widrow}}]{2013Natur.493...62I}
{Ibata}, R.~A., {Lewis}, G.~F., {Conn}, A.~R., {et~al.} 2013, NAT, 493, 62

\bibitem[{{Jerjen} {et~al.}(2000{\natexlab{a}}){Jerjen}, {Binggeli}, \&
  {Freeman}}]{2000AJ....119..593J}
{Jerjen}, H., {Binggeli}, B., \& {Freeman}, K.~C. 2000{\natexlab{a}}, \aj, 119,
  593

\bibitem[{{Jerjen} {et~al.}(2000{\natexlab{b}}){Jerjen}, {Freeman}, \&
  {Binggeli}}]{2000AJ....119..166J}
{Jerjen}, H., {Freeman}, K.~C., \& {Binggeli}, B. 2000{\natexlab{b}}, \aj, 119,
  166

\bibitem[{{Karachentsev} {et~al.}(2004){Karachentsev}, {Karachentseva},
  {Huchtmeier}, \& {Makarov}}]{2004AJ....127.2031K}
{Karachentsev}, I.~D., {Karachentseva}, V.~E., {Huchtmeier}, W.~K., \&
  {Makarov}, D.~I. 2004, \aj, 127, 2031

\bibitem[{{Karachentsev} {et~al.}(2013){Karachentsev}, {Makarov}, \&
  {Kaisina}}]{2013AJ....145..101K}
{Karachentsev}, I.~D., {Makarov}, D.~I., \& {Kaisina}, E.~I. 2013, \aj, 145,
  101

\bibitem[{{Karachentsev} {et~al.}(2007){Karachentsev}, {Tully}, {Dolphin},
  {Sharina}, {Makarova}, {Makarov}, {Sakai}, {Shaya}, {Kashibadze},
  {Karachentseva}, \& {Rizzi}}]{2007AJ....133..504K}
{Karachentsev}, I.~D., {Tully}, R.~B., {Dolphin}, A., {et~al.} 2007, \aj, 133,
  504

\bibitem[{{Kirby} {et~al.}(2012){Kirby}, {Koribalski}, {Jerjen}, \&
  {L{\'o}pez-S{\'a}nchez}}]{2012MNRAS.420.2924K}
{Kirby}, E.~M., {Koribalski}, B., {Jerjen}, H., \& {L{\'o}pez-S{\'a}nchez},
  {\'A}. 2012, \mnras, 420, 2924

\bibitem[{{Koch} \& {Grebel}(2006)}]{2006AJ....131.1405K}
{Koch}, A. \& {Grebel}, E.~K. 2006, \aj, 131, 1405

\bibitem[{{Koribalski} {et~al.}(2004){Koribalski}, {Staveley-Smith}, {Kilborn},
  {Ryder}, {Kraan-Korteweg}, {Ryan-Weber}, {Ekers}, {Jerjen}, {Henning},
  {Putman}, {Zwaan}, {de Blok}, {Calabretta}, {Disney}, {Minchin}, {Bhathal},
  {Boyce}, {Drinkwater}, {Freeman}, {Gibson}, {Green}, {Haynes}, {Juraszek},
  {Kesteven}, {Knezek}, {Mader}, {Marquarding}, {Meyer}, {Mould}, {Oosterloo},
  {O'Brien}, {Price}, {Sadler}, {Schr{\"o}der}, {Stewart}, {Stootman}, {Waugh},
  {Warren}, {Webster}, \& {Wright}}]{2004AJ....128...16K}
{Koribalski}, B.~S., {Staveley-Smith}, L., {Kilborn}, V.~A., {et~al.} 2004,
  \aj, 128, 16

\bibitem[{{Kroupa} {et~al.}(2005){Kroupa}, {Theis}, \&
  {Boily}}]{2005A&A...431..517K}
{Kroupa}, P., {Theis}, C., \& {Boily}, C.~M. 2005, \aap, 431, 517

\bibitem[{{Kunkel} \& {Demers}(1976)}]{1976RGOB..182..241K}
{Kunkel}, W.~E. \& {Demers}, S. 1976, in Royal Greenwich Observatory Bulletins,
  Vol. 182, The Galaxy and the Local Group, ed. R.~J. {Dickens}, J.~E. {Perry},
  F.~G. {Smith}, \& I.~R. {King}, 241

\bibitem[{{Libeskind} {et~al.}(2019){Libeskind}, {Carlesi}, {M{\"u}ller},
  {Pawlowski}, {Hoffman}, {Pomar{\`e}de}, {Courtois}, {Tully}, {Gottl{\"o}ber},
  {Steinmetz}, {Sorce}, \& {Knebe}}]{2019MNRAS.490.3786L}
{Libeskind}, N.~I., {Carlesi}, E., {M{\"u}ller}, O., {et~al.} 2019, \mnras,
  490, 3786

\bibitem[{{Libeskind} {et~al.}(2005){Libeskind}, {Frenk}, {Cole}, {Helly},
  {Jenkins}, {Navarro}, \& {Power}}]{2005MNRAS.363..146L}
{Libeskind}, N.~I., {Frenk}, C.~S., {Cole}, S., {et~al.} 2005, \mnras, 363, 146

\bibitem[{{Libeskind} {et~al.}(2009){Libeskind}, {Frenk}, {Cole}, {Jenkins}, \&
  {Helly}}]{2009MNRAS.399..550L}
{Libeskind}, N.~I., {Frenk}, C.~S., {Cole}, S., {Jenkins}, A., \& {Helly},
  J.~C. 2009, \mnras, 399, 550

\bibitem[{{Libeskind} {et~al.}(2016){Libeskind}, {Guo}, {Tempel}, \&
  {Ibata}}]{2016ApJ...830..121L}
{Libeskind}, N.~I., {Guo}, Q., {Tempel}, E., \& {Ibata}, R. 2016, \apj, 830,
  121

\bibitem[{{Libeskind} {et~al.}(2015){Libeskind}, {Hoffman}, {Tully},
  {Courtois}, {Pomar{\`e}de}, {Gottl{\"o}ber}, \&
  {Steinmetz}}]{2015MNRAS.452.1052L}
{Libeskind}, N.~I., {Hoffman}, Y., {Tully}, R.~B., {et~al.} 2015, \mnras, 452,
  1052

\bibitem[{{Libeskind} {et~al.}(2010){Libeskind}, {Yepes}, {Knebe},
  {Gottl{\"o}ber}, {Hoffman}, \& {Knollmann}}]{2010MNRAS.401.1889L}
{Libeskind}, N.~I., {Yepes}, G., {Knebe}, A., {et~al.} 2010, \mnras, 401, 1889

\bibitem[{{Lipnicky} \& {Chakrabarti}(2017)}]{2017MNRAS.468.1671L}
{Lipnicky}, A. \& {Chakrabarti}, S. 2017, \mnras, 468, 1671

\bibitem[{{Lynden-Bell}(1976)}]{1976MNRAS.174..695L}
{Lynden-Bell}, D. 1976, \mnras, 174, 695

\bibitem[{{Lynden-Bell}(1982)}]{1982Obs...102..202L}
{Lynden-Bell}, D. 1982, The Observatory, 102, 202

\bibitem[{{McConnachie} \& {Irwin}(2006)}]{2006MNRAS.365..902M}
{McConnachie}, A.~W. \& {Irwin}, M.~J. 2006, \mnras, 365, 902

\bibitem[{{Metz} {et~al.}(2008){Metz}, {Kroupa}, \&
  {Libeskind}}]{2008ApJ...680..287M}
{Metz}, M., {Kroupa}, P., \& {Libeskind}, N.~I. 2008, \apj, 680, 287

\bibitem[{{M{\"u}ller} {et~al.}(2020){M{\"u}ller}, {Fahrion}, {Rejkuba},
  {Hilker}, {Lelli}, {Lutz}, {Pawlowski}, {Coccato}, {Anand}, \&
  {Jerjen}}]{2020arXiv201104990M}
{M{\"u}ller}, O., {Fahrion}, K., {Rejkuba}, M., {et~al.} 2020, arXiv e-prints,
  arXiv:2011.04990

\bibitem[{{M{\"u}ller} {et~al.}(2015){M{\"u}ller}, {Jerjen}, \&
  {Binggeli}}]{2015A&A...583A..79M}
{M{\"u}ller}, O., {Jerjen}, H., \& {Binggeli}, B. 2015, \aap, 583, A79

\bibitem[{{M{\"u}ller} {et~al.}(2017{\natexlab{a}}){M{\"u}ller}, {Jerjen}, \&
  {Binggeli}}]{2017A&A...597A...7M}
{M{\"u}ller}, O., {Jerjen}, H., \& {Binggeli}, B. 2017{\natexlab{a}}, \aap,
  597, A7

\bibitem[{{M{\"u}ller} {et~al.}(2016){M{\"u}ller}, {Jerjen}, {Pawlowski}, \&
  {Binggeli}}]{2016A&A...595A.119M}
{M{\"u}ller}, O., {Jerjen}, H., {Pawlowski}, M.~S., \& {Binggeli}, B. 2016,
  \aap, 595, A119

\bibitem[{{M{\"u}ller} {et~al.}(2018{\natexlab{a}}){M{\"u}ller}, {Pawlowski},
  {Jerjen}, \& {Lelli}}]{2018Sci...359..534M}
{M{\"u}ller}, O., {Pawlowski}, M.~S., {Jerjen}, H., \& {Lelli}, F.
  2018{\natexlab{a}}, Science, 359, 534

\bibitem[{{M{\"u}ller} {et~al.}(2018{\natexlab{b}}){M{\"u}ller}, {Rejkuba}, \&
  {Jerjen}}]{MuellerTRGB2018}
{M{\"u}ller}, O., {Rejkuba}, M., \& {Jerjen}, H. 2018{\natexlab{b}}, \aap, 615,
  A96

\bibitem[{{M{\"u}ller} {et~al.}(2019){M{\"u}ller}, {Rejkuba}, {Pawlowski},
  {Ibata}, {Lelli}, {Hilker}, \& {Jerjen}}]{MuellerTRGB2019}
{M{\"u}ller}, O., {Rejkuba}, M., {Pawlowski}, M.~S., {et~al.} 2019, \aap, 629,
  A18

\bibitem[{{M{\"u}ller} {et~al.}(2017{\natexlab{b}}){M{\"u}ller}, {Scalera},
  {Binggeli}, \& {Jerjen}}]{2017A&A...602A.119M}
{M{\"u}ller}, O., {Scalera}, R., {Binggeli}, B., \& {Jerjen}, H.
  2017{\natexlab{b}}, \aap, 602, A119

\bibitem[{{Nelson} {et~al.}(2019){Nelson}, {Springel}, {Pillepich},
  {Rodriguez-Gomez}, {Torrey}, {Genel}, {Vogelsberger}, {Pakmor}, {Marinacci},
  {Weinberger}, {Kelley}, {Lovell}, {Diemer}, \&
  {Hernquist}}]{2019ComAC...6....2N}
{Nelson}, D., {Springel}, V., {Pillepich}, A., {et~al.} 2019, Computational
  Astrophysics and Cosmology, 6, 2

\bibitem[{{Pawlowski}(2018)}]{2018MPLA...3330004P}
{Pawlowski}, M.~S. 2018, Modern Physics Letters A, 33, 1830004

\bibitem[{{Pawlowski} {et~al.}(2019){Pawlowski}, {Bullock}, {Kelley}, \&
  {Famaey}}]{2019ApJ...875..105P}
{Pawlowski}, M.~S., {Bullock}, J.~S., {Kelley}, T., \& {Famaey}, B. 2019, \apj,
  875, 105

\bibitem[{{Pawlowski} {et~al.}(2015){Pawlowski}, {Famaey}, {Merritt}, \&
  {Kroupa}}]{2015ApJ...815...19P}
{Pawlowski}, M.~S., {Famaey}, B., {Merritt}, D., \& {Kroupa}, P. 2015, \apj,
  815, 19

\bibitem[{{Pawlowski} \& {Kroupa}(2020)}]{2020MNRAS.491.3042P}
{Pawlowski}, M.~S. \& {Kroupa}, P. 2020, \mnras, 491, 3042

\bibitem[{{Pawlowski} {et~al.}(2012){Pawlowski}, {Pflamm-Altenburg}, \&
  {Kroupa}}]{2012MNRAS.423.1109P}
{Pawlowski}, M.~S., {Pflamm-Altenburg}, J., \& {Kroupa}, P. 2012, \mnras, 423,
  1109

\bibitem[{{Peterson} \& {Caldwell}(1993)}]{1993AJ....105.1411P}
{Peterson}, R.~C. \& {Caldwell}, N. 1993, \aj, 105, 1411

\bibitem[{{Phillips} {et~al.}(2015){Phillips}, {Cooper}, {Bullock}, \&
  {Boylan-Kolchin}}]{2015MNRAS.453.3839P}
{Phillips}, J.~I., {Cooper}, M.~C., {Bullock}, J.~S., \& {Boylan-Kolchin}, M.
  2015, \mnras, 453, 3839

\bibitem[{{Pillepich} {et~al.}(2018){Pillepich}, {Nelson}, {Hernquist},
  {Springel}, {Pakmor}, {Torrey}, {Weinberger}, {Genel}, {Naiman}, {Marinacci},
  \& {Vogelsberger}}]{2018MNRAS.475..648P}
{Pillepich}, A., {Nelson}, D., {Hernquist}, L., {et~al.} 2018, \mnras, 475, 648

\bibitem[{{Planck Collaboration} {et~al.}(2016){Planck Collaboration}, {Ade},
  {Aghanim}, {Arnaud}, {Ashdown}, {Aumont}, {Baccigalupi}, {Banday},
  {Barreiro}, {Bartlett}, \& et~al.}]{2016A&A...594A..13P}
{Planck Collaboration}, {Ade}, P.~A.~R., {Aghanim}, N., {et~al.} 2016, \aap,
  594, A13

\bibitem[{{Puzia} \& {Sharina}(2008)}]{2008ApJ...674..909P}
{Puzia}, T.~H. \& {Sharina}, M.~E. 2008, \apj, 674, 909

\bibitem[{{Samuel} {et~al.}(2020){Samuel}, {Wetzel}, {Chapman}, {Tollerud},
  {Hopkins}, {Boylan-Kolchin}, {Bailin}, \&
  {Faucher-Gigu{\`e}re}}]{2020arXiv201008571S}
{Samuel}, J., {Wetzel}, A., {Chapman}, S., {et~al.} 2020, arXiv e-prints,
  arXiv:2010.08571

\bibitem[{{Santos-Santos} {et~al.}(2020){Santos-Santos}, {Dominguez-Tenreiro},
  {Artal}, {Pedrosa}, {Bignone}, {Martinez-Serrano}, {Gomez-Flechoso},
  {Tissera}, \& {Serna}}]{2020arXiv200411585S}
{Santos-Santos}, I., {Dominguez-Tenreiro}, R., {Artal}, H., {et~al.} 2020,
  arXiv e-prints, arXiv:2004.11585

\bibitem[{{Saviane} \& {Jerjen}(2007)}]{2007AJ....133.1756S}
{Saviane}, I. \& {Jerjen}, H. 2007, \aj, 133, 1756

\bibitem[{{Sawala} {et~al.}(2016){Sawala}, {Frenk}, {Fattahi}, {Navarro},
  {Bower}, {Crain}, {Dalla Vecchia}, {Furlong}, {Helly}, {Jenkins}, {Oman},
  {Schaller}, {Schaye}, {Theuns}, {Trayford}, \& {White}}]{2016MNRAS.457.1931S}
{Sawala}, T., {Frenk}, C.~S., {Fattahi}, A., {et~al.} 2016, \mnras, 457, 1931

\bibitem[{Smith {et~al.}(2016)Smith, Duc, Bournaud, \& Yi}]{Smith2016}
Smith, R., Duc, P.~A., Bournaud, F., \& Yi, S.~K. 2016, \apj, 818, 11

\bibitem[{{Sohn} {et~al.}(2017){Sohn}, {Patel}, {Besla}, {van der Marel},
  {Bullock}, {Strigari}, {van de Ven}, {Walker}, \&
  {Bellini}}]{2017ApJ...849...93S}
{Sohn}, S.~T., {Patel}, E., {Besla}, G., {et~al.} 2017, \apj, 849, 93

\bibitem[{{Sohn} {et~al.}(2020){Sohn}, {Patel}, {Fardal}, {Besla}, {van der
  Marel}, {Geha}, \& {Guhathakurta}}]{2020arXiv200806055S}
{Sohn}, S.~T., {Patel}, E., {Fardal}, M.~A., {et~al.} 2020, arXiv e-prints,
  arXiv:2008.06055

\bibitem[{{Springel} {et~al.}(2018){Springel}, {Pakmor}, {Pillepich},
  {Weinberger}, {Nelson}, {Hernquist}, {Vogelsberger}, {Genel}, {Torrey},
  {Marinacci}, \& {Naiman}}]{2018MNRAS.475..676S}
{Springel}, V., {Pakmor}, R., {Pillepich}, A., {et~al.} 2018, \mnras, 475, 676

\bibitem[{{Springel} {et~al.}(2005){Springel}, {White}, {Jenkins}, {Frenk},
  {Yoshida}, {Gao}, {Navarro}, {Thacker}, {Croton}, {Helly}, {Peacock}, {Cole},
  {Thomas}, {Couchman}, {Evrard}, {Colberg}, \& {Pearce}}]{2005Natur.435..629S}
{Springel}, V., {White}, S.~D.~M., {Jenkins}, A., {et~al.} 2005, \nat, 435, 629

\bibitem[{{Taylor} {et~al.}(2018){Taylor}, {Eigenthaler}, {Puzia}, {Mu{\~n}oz},
  {Ribbeck}, {Zhang}, {Ordenes-Brice{\~n}o}, \& {Bovill}}]{2018ApJ...867L..15T}
{Taylor}, M.~A., {Eigenthaler}, P., {Puzia}, T.~H., {et~al.} 2018, \apjl, 867,
  L15

\bibitem[{{Tully} {et~al.}(2015){Tully}, {Libeskind}, {Karachentsev},
  {Karachentseva}, {Rizzi}, \& {Shaya}}]{2015ApJ...802L..25T}
{Tully}, R.~B., {Libeskind}, N.~I., {Karachentsev}, I.~D., {et~al.} 2015,
  \apjl, 802, L25

\bibitem[{{Tully} {et~al.}(2008){Tully}, {Shaya}, {Karachentsev}, {Courtois},
  {Kocevski}, {Rizzi}, \& {Peel}}]{2008ApJ...676..184T}
{Tully}, R.~B., {Shaya}, E.~J., {Karachentsev}, I.~D., {et~al.} 2008, \apj,
  676, 184

\bibitem[{{Vogelsberger} {et~al.}(2014){Vogelsberger}, {Genel}, {Springel},
  {Torrey}, {Sijacki}, {Xu}, {Snyder}, {Nelson}, \&
  {Hernquist}}]{2014MNRAS.444.1518V}
{Vogelsberger}, M., {Genel}, S., {Springel}, V., {et~al.} 2014, \mnras, 444,
  1518

\bibitem[{{Wang} {et~al.}(2020){Wang}, {Hammer}, {Rejkuba}, {Crnojevi{\'c}}, \&
  {Yang}}]{Wang2020}
{Wang}, J., {Hammer}, F., {Rejkuba}, M., {Crnojevi{\'c}}, D., \& {Yang}, Y.
  2020, \mnras, 498, 2766

\bibitem[{{Zentner} {et~al.}(2005){Zentner}, {Kravtsov}, {Gnedin}, \&
  {Klypin}}]{2005ApJ...629..219Z}
{Zentner}, A.~R., {Kravtsov}, A.~V., {Gnedin}, O.~Y., \& {Klypin}, A.~A. 2005,
  \apj, 629, 219

\end{thebibliography}

\end{document}